%% file: VectorcalculusEd.tex
\RequirePackage{mmap} 
\documentclass[aps,reprint,superscriptaddress,prstper,showpacs]{revtex4-1}
\usepackage{amsfonts}
\usepackage{amsmath}
\usepackage{amssymb}
\usepackage[usenames,dvipsnames,table]{xcolor}
\usepackage{graphicx}
\usepackage{epstopdf}
\usepackage{float}
\usepackage{enumitem}
\usepackage{array}
\usepackage{multirow}
\usepackage{hyperref}
\hypersetup{colorlinks=true, linkcolor=PineGreen ,citecolor=BrickRed , urlcolor=Blue}
\renewcommand{\vec}[1]{\boldsymbol{\mathbf{#1}}}
\let\oldhat\hat
\renewcommand{\hat}[1]{\, \boldsymbol{\oldhat{\mathbf{#1}}}}

\begin{document}

\date{\today}
\title{Students' difficulties with vector calculus in electrodynamics}
\author{Laurens Bollen}\email{Laurens.Bollen@fys.kuleuven.be}\affiliation{Department of Physics and Astronomy \& LESEC, KU Leuven, Celestijnenlaan 200c, 3001 Leuven, Belgium.}
\author{Paul van Kampen}\email{Paul.van.Kampen@dcu.ie}\affiliation{Centre for the Advancement of Science and Mathematics Teaching and Learning \& School of Physical Sciences, Dublin City University, Glasnevin, Dublin 9, Ireland.}
\author{Mieke De Cock}\email{Mieke.DeCock@fys.kuleuven.be}\affiliation{Department of Physics and Astronomy \& LESEC, KU Leuven, Celestijnenlaan 200c, 3001 Leuven, Belgium.}

\begin{abstract}
Understanding Maxwell's equations in differential form is of great importance when studying the electrodynamic phenomena discussed in advanced electromagnetism courses. It is therefore necessary that students master the use of vector calculus in physical situations. In this light we investigated the difficulties second year students at KU Leuven encounter with the divergence and curl of a vector field in mathematical and physical contexts. We have found that they are quite skilled at doing calculations, but struggle with interpreting graphical representations of vector fields and applying vector calculus to physical situations. We have found strong indications that traditional instruction is not sufficient for our students to fully understand the meaning and power of Maxwell's equations in electrodynamics.
\end{abstract}


\maketitle

\section{Introduction}
It is difficult to overestimate the importance of Maxwell's equations in the study of electricity and magnetism. Together with the Lorentz force law, these equations provide the foundations of classical electrodynamics. They can be used to solve problems concerning electromagnetic phenomena including those occurring in electric circuits and wave optics. Moreover, Maxwell's equations are the first example of a gauge theory (which is commonly used in particle physics) and are the starting point for Einstein's theory of special relativity, both historically and in many curricula. It is therefore desirable that students have a profound understanding of these equations.\par  

For physics education researchers, an electrodynamics course is an ideal `laboratory' to explore the link between mathematics and physics since students have acquired knowledge of the physical concepts in an introductory electricity and magnetism course, and have learned the necessary mathematical techniques during instruction on calculus. To describe more complex electromagnetic phenomena, they will need to apply this advanced mathematics in the description of the physical reality. This study has charted some problems our students encounter with physics and mathematics when using Maxwell's equations.\par

Maxwell's equations can be formulated in differential or in integral form. In differential form, the four laws are written in the language of vector calculus that includes the differential operators divergence and curl. These are typically expressed using the nabla $\boldsymbol{\nabla}$ symbol to denote the del operator. The divergence of a vector field ($\boldsymbol{\nabla} \cdot \vec{A}$) is a scalar quantity that measures the magnitude of a source or sink of the field at a given point. The curl of a vector field ($\boldsymbol{\nabla} \times \vec{A}$) results in a vector field that describes the infinitesimal rotation at any point in the field. Both quantities are defined locally: they only describe the characteristics of a vector field at a single point. This is the most important distinction from Maxwell's equations in integral form, which describe the electromagnetic field in a region of space. For physics majors, the two formulations are equally important since they both have certain advantages and limitations in specific contexts. The research described in this paper focuses on students' understanding of Maxwell's equations in differential form and includes their knowledge and skills concerning vector calculus.\par

In Section~\ref{sec:literature}\ we provide an overview of the related literature, including work on the link between mathematics and physics: research on electromagnetism, vector calculus and the combination of the two. This leads to our research goals, which are described in Section~\ref{sec:design}. To formulate our research questions, we distinguish four kinds of skills that need to be acquired by the students:
\begin{enumerate}
  \item structural understanding\cite{Sfard1991,Tuminaro2004a}\ of the concepts of gradient, divergence and curl
  \item the interpretation of these operators in the context of a graphical representation of the field
  \item doing calculations that involve vector operators
  \item conceptual understanding of Maxwell's equations in differential form. 
\end{enumerate}
The educational context and methodology are described in Section~\ref{sec:context}, followed by the results of our study in Section~\ref{sec:results}. This section contains a discussion of our students' skills and difficulties concerning the four topics that are listed above. The most important findings and possible implications for teaching are summarized in Section~\ref{sec:conclusions}.

\section{Related literature}\label{sec:literature}
The majority of physics education research (PER) at university level to date concerns introductory courses (examples can be found in the summary of McDermott and Redish\cite{McDermott1999}). These studies have, among other things, yielded an extensive inventory of conceptual problems in physics and the finding that many students struggle with the application of their mathematical knowledge in a physical context.\cite{Artigue1990,Yeatts1992,Redish2006}\ The role of mathematics in physics education has been an important topic in recent PER projects.\cite{Tuminaro2004,Redish2006,Bing2007,Tuminaro2007,Bing2009,Wilcox2013,Karam2014}\ Manogue and Dray went so far as to state that physicists and mathematicians speak a different language, but use the same vocabulary.\cite{Manogue2004}\  A tendency for students to focus on equations and calculations rather than on the physical meaning behind the symbols has been identified as a recurring issue.\cite{Larkin1980,Sherin2001,Kuo2013}\ Another source of difficulties is the use of multiple representations in physics and mathematics. Students have severe difficulties combining the information in texts, equations, symbols, graphs and figures into a single unambiguous story.\cite{Kohl2005,Kohl2006,Kohl2008,Nguyen2009,DeCock2012,Wagner2012}\ Our work adds to the investigation of these issues in the context of an intermediate electrodynamics course.\par

Introductory electricity and magnetism (E\&M) courses are a popular setting to learn about student misconceptions.\cite{Maloney2001,Maloney1985,Adrian1997,Albe2001,Ferguson-Hessler1987,Guisasola2004,Kesonen2011,Smaill2012,Garzon2014,Guisasola2010,Preyer2000,Sherwood2009,Sihvola2012}\ A fair amount of research has been carried out on the use of integrals in E\&M.\cite{Doughty2014,Guisasola2008,Manogue2006,Pepper2010,Singh2005,Traxler2007,Wallace2010,Nguyen2011,Hu2013}\ This research informs the work we present here, since we have adopted some of the ideas and methodologies in these papers as a starting point for our own research. At Dublin City University students' ideas about integrals were investigated using an approach based on the idea of the concept image, i.e. all the mental processes activated when students encounter a certain concept (e.g. an integral of a vector operator).\cite{Doughty2014}\ It is unique for every person, and therefore differs from the (formal) concept definition, which is a description that is accepted by the wider community.\cite{Tall1981}\ One aspect of our study concerns our students' concept image of vector operators, which we relate to the difficulties they encounter when applying Maxwell's equations in differential form.\par

In advanced courses (often called E\&M2 or electrodynamics) vector calculus plays an important role. It is known that operations with vectors and vector fields (e.g. vector addition and the dot product) provide students with a many problems.\cite{Barniol2012,Doughty2013,Knight1995,Nguyen2003,Deventer2008,Barniol2014}\ Furthermore students struggle with the use of vectors in different coordinate systems and the application of appropriate unit vectors.\cite{Dray2003,Hinrichs2010,Barniol2012,Barniol2014}\ However, little is known about situations where the del operator is applied to scalar or vector fields. Gire and Price discussed the option to use graphical representations when teaching about vector fields and vector calculus. Based on their experience with different types of in-class activities, they argued that algebraic representations are useful since they can easily be manipulated, but students gain more insight into the differences between components and coordinates when using a graphical approach. Moreover they expect that students will benefit from being able to translate one representation to another.\cite{Gire2012}\ Singh and Maries report that about half of their graduate students before instruction, and one out of three after instruction, are unable to determine where the divergence or curl is (non)zero when provided with a graphical representation of a vector field. They argue that physics courses are often a missed learning opportunity because they strongly focus on mathematics but fail to develop a functional understanding of the underlying concepts.\cite{Singh2013}\par

While there is quite some physics and mathematics education research on vector calculus, the amount of research on vector calculus in the context of electrodynamics is limited. The educational setting in this context however is different. Manogue and Dray pointed out that in mathematics the gradient, divergence and curl are used in a general and abstract way, while in physics they are mostly used in certain symmetries (Cartesian, cylindrical or spherical).\cite{Manogue1998}\ Research at the University of Colorado showed that problems arise when asking students to determine where the divergence of an electric field vanishes for a given charge distribution. This type of question can be solved with the differential form of Gauss's law in a straightforward way. However, the authors report that only 26\% of their students were able to give a correct answer.\cite{Pepper2012}\ Baily and Astolfi found that what students from St Andrews learned about the divergence in one context (Gauss's law) often did not translate to their understanding in other contexts (e.g. the continuity equation).\cite{Baily2014} \par

In summary, the literature reviewed here shows that students struggle when they have to use their knowledge from mathematics in a physical context. Clearly, this also applies to the specific case of applying vector calculus in electrodynamics. However, there is still a lot of research to be done on the subject. In the next section, the contribution of our study is formulated in terms of goals and research questions.

\section{Research design}\label{sec:design}
This paper gives an account of an exploratory study of students' strengths and weaknesses in using vector calculus in mathematical and physical contexts. The research extends the previous findings mentioned in Section~\ref{sec:literature}\ by adopting a broader approach to ascertain the knowledge, skills and understanding our students have acquired.  The goal of this study is twofold: it aims to provide both researchers and teachers with insights into the learning results of traditional instruction of electrodynamics, and it is the first stage of a large scale investigation of students' understanding of Maxwell's equations in differential form.\par

In this first stage of the investigation we aim to gain insight into the difficulties students encounter with vector calculus in a purely mathematical or physical context. To this end we distinguish four different kinds of skills and competencies students need to acquire: structural understanding~\cite{Sfard1991,Tuminaro2004a}\ of divergence and curl, graphical interpretation of vector fields, calculation of divergence and curl, and conceptual understanding of Maxwell's equations in differential form. Based on their instruction prior to the electrodynamics course and results from the literature, we expect our students to have reasonable facility with the mathematical techniques needed to carry out calculations while lacking experience with interpreting graphical representations of vector fields. We have investigated our students' attainment at the start of the electrodynamics course and have tried to establish to what extent their understanding of Maxwell's equations in differential form changes while taking the course. We have focused on the following research questions:
\begin{itemize}
\item Did our students acquire a structural understanding~\cite{Sfard1991,Tuminaro2004a}\ of gradient, divergence and curl from their introductory and intermediate mathematics courses?
\begin{itemize}
\item What is their concept image~\cite{Tall1981}\ of gradient, divergence and curl?
\item How do they describe the meaning of the vector operators?
\end{itemize}
\item Can students interpret a graphical representation of a vector field in terms of its divergence and curl?
\begin{itemize}
\item Can they deduce where the divergence and curl of vector fields are (non)zero in a purely mathematical context?
\item Can they deduce where the divergence and curl of electromagnetic fields are (non)zero?
\item What strategies do they use to interpret these representations?
\end{itemize}
\item Did our students acquire the necessary mathematical techniques to perform calculations involving vector operators with and without a physical context?
\begin{itemize}
\item What technical difficulties do they encounter?
\item Do different kinds of coordinate systems (Cartesian, cylindrical, spherical) present different challenges?
\end{itemize}
\item Do students conceptually understand Maxwell's equations in differential form?
\begin{itemize}
\item Are they able to correctly deduce whether the divergence and curl of an electromagnetic field are zero or non-zero in a given situation?
\end{itemize}
\end{itemize}

\section{Educational context and methodology}\label{sec:context}
To answer the research questions, we gave written paper-and-pencil questions to second year university students in a traditional thirteen week intermediate electrodynamics course. The students major in physics or mathematics at the KU Leuven. They use Griffiths' textbook~\cite{Griffiths2012}\ and are instructed in one two-hour lecture and one two-hour problem solving session per week in which they discuss typical end of chapter problems from the textbook. The students have already completed an introductory electromagnetism course using the textbook of Serway and Jewett~\cite{Serway2009}\ that leads up to Maxwell's equations in integral form, and at least two calculus courses\footnote{An introductory calculus course based on the textbook by Adams and Essex,\cite{Adams2009} and a course on differential equations including some paragraphs, examples and exercises about gradient, divergence and curl}\ that include a chapter on vector calculus. Therefore they have encountered the necessary mathematical tools and physical situations presented in the electrodynamics course.\par

To identify the prior knowledge of our students they were given a pretest before the first lecture in the advanced electromagnetism course based on Griffiths' textbook.~\cite{Griffiths2012}\ Since these students had encountered vector calculus mostly in a mathematical setting, the questions on the pretest do not contain any physical context. To encourage students to write down their reasoning, calculations and thinking process, all questions were open-ended. A post-test was given after instruction on chapters 1--7 of Griffiths' textbook \cite{Griffiths2012}, during a lecture about halfway through the semester. The post-test assignments are mostly similar to those on the pretest; however, a physical context is introduced in some cases to investigate whether information on the physical situation affects the students' ability to interpret the divergence and curl of the (electromagnetic) vector fields. The post-test also comprises questions that evaluate students' understanding of Maxwell's equations. There were no time constraints for the students to complete the pre- and post-test.\par

The analysis focuses on the solution method and thinking process rather than the result. To describe and explain the variation in students' conceptions, ideas of phenomenography are used. Phenomenography is an empirical approach that aims to identify and categorize the different qualitative ways in which different people perceive and understand phenomena.\cite{Guisasola2004,Marton1981}\ The categories used in the analysis of our data were established in a bottom-up approach where one of the authors proposed a set of categories based on the answers students gave, the strategies they used and the mistakes they made. After an elaborate discussion with the other collaborators about some specific student answers, we refined our classification and decided on a final set of categories. To confirm that our categories are well defined, we evaluated the inter-rater reliability by calculating Cohen's kappa ($\kappa$). For individual questions, Cohen's kappa ranged from 0.76 to 1.00, indicating a \emph{substantial}\ to \emph{almost perfect}\ agreement. Since the number of students is limited ($N=30$ on the pretest and $N=19$ on the post-test), the percentages should be generalized with care. Nevertheless they should give a clear view of the limitations in our students' understanding of vector calculus in electrodynamics.

\section{Results and discussion}\label{sec:results}
In this section the results of the pre- and post-test are presented and discussed. The questions can be found in Appendices~\ref{app:pretest}\ and~\ref{app:posttest}.

\subsection{Pretest}
The pretest shown in Appendix~\ref{app:pretest}\ was given to all 30 students at the beginning of the course to probe their knowledge and understanding acquired in previous courses. The first part of the pretest identifies students' concept images~\cite{Tall1981}\ of the operators grad, div and curl. In the second part, the students' calculational skills and their ability to interpret graphical representations of vector fields are tested. For this part only, some useful formulas were attached to the questions (Appendix ~\ref{app:formulas}).\par

The three questions on the pretest correspond to the first three research questions that were discussed above. Since our students only studied Maxwell's equations in integral form during their introductory course, we did not include a question that assesses their understanding of the differential form.

\subsubsection{Concept image of grad, div and curl}
The concept image question serves to get a better understanding of what students associate with the gradient, divergence and curl in a very general sense. Expressions for grad, div and curl are given to the students, and they are asked to write down everything they think of. From this, we can make some statements about the students' concept image~\cite{Tall1981,Vinner1989,Doughty2014}\  of these operators. The students' responses to the questions are described qualitatively in Table~\ref{tab:conceptimage}. We distinguished three important emerging categories in the students' answers: information about the structural meaning of the vector operators, the scalar or vector character of the expression, and the name and symbolic expression that students wrote down. Obviously, a student can give more than one interpretation and therefore the percentages sum to more than 100\%. We do not suggest that students do not know something they did not write, but we do think the question reveals what is cued first and foremost.\par

\begin{table}[htbp]
  \centering
  \caption{Categorization of students' interpretation of the expressions $\boldsymbol{\nabla} A$, $\boldsymbol{\nabla} \cdot \vec{A}$ and $\boldsymbol{\nabla} \times \vec{A}$.}
    \begin{ruledtabular}
		\begin{tabular}{lccc}
		   \multicolumn{1}{c}{Category $(N=30)$} & $\boldsymbol{\nabla} A$ & $\boldsymbol{\nabla} \cdot \vec{A}$ & $\boldsymbol{\nabla} \times \vec{A}$\\
		\hline
    Correct concept & 10\% & 0\% & 10\% \\
    Incorrect/incomplete concept & 23\% & 10\% & 7\% \\
    \rule{0pt}{3ex}Scalar & 3\% & 63\% & 0\% \\
    Vector & 60\% & 7\% & 53\% \\
    \rule{0pt}{3ex}Naming & 70\% & 50\% & 53\% \\
    Formula & 63\% & 50\% & 37\% \\
    \rule{0pt}{3ex}Other & 3\% & 3\% & 7\% \\
    \rule{0pt}{3ex}No answer & 3\% & 10\% & 10\% \\
    \end{tabular}%
		\end{ruledtabular}
  \label{tab:conceptimage}%
\end{table}%

Only a few students gave a description of the operators we deemed conceptual. Some provided a more or less correct description that resembles the concept definition:
\begin{quote}\textit{ ``The curl tells you how strong and which way the vector field $\vec{A}$ rotates.'' }\end{quote}
This student did not make a statement about the local character of the curl, but does seem to have a rather good idea about what the curl represents. Others had very incomplete or incorrect conceptual ideas: 
\begin{quote}\textit{ ``The divergence is a measure for how the field is changing.'' }\end{quote} \begin{quote}\textit{ ``The gradient of $A$ is the vector normal to the plane.'' }\end{quote}\par
We also observed that some students misidentified the vector or scalar character of the expression. More than half of the students mentioned which operations result in a vector field, and which produce a scalar field. Sometimes they explicitly wrote it down, in other cases it could be derived from their notation. The notation in this student's answer for example shows he thinks the divergence of a vector field is a vector: 
\begin{quote}\textit{``In three dimensions this is the divergence and therefore \small{$\overrightarrow{\nabla} \cdot \overrightarrow{A} = \frac{\partial\overrightarrow{A}}{\partial x} + \frac{\partial\overrightarrow{A}}{\partial y} + \frac{\partial\overrightarrow{A}}{\partial z}$''}}\end{quote}
One of our students wrote that the gradient of a scalar is again a scalar. Two students also seemed to think the divergence of a vector field is again a vector field.  We did not observe a single misidentification of the vector character of the curl of a vector field. This corroborates the findings of Barniol and Zavala, who showed that students have significantly more problems with the vector or scalar nature of the dot product than of the vector product.\cite{Barniol2014}\ Since the students correctly described $A$ as being a scalar (field) and $\vec{A}$ as being a vector (field), we have no indications that there was a problem with the notation in the question. About one out of three students did not make any statement concerning the vector or scalar character in their answer (e.g. they just named $\boldsymbol{\nabla} \cdot \vec{A}$ ``divergence'', without any explanation).\par

The category `Naming' contains all students who wrote down the correct name of the expression. No students remembered names incorrectly or mixed up the terms gradient, divergence and curl. Nevertheless, half of the students did not explicitly identify $\boldsymbol{\nabla} \cdot \vec{A}$ as the divergence and $\boldsymbol{\nabla} \times\vec{A}$ as the curl of a vector field. Only five of our students wrote that $\boldsymbol{\nabla}$ is the nabla symbol, and one student called it the del symbol. About one out of four however called this symbol the gradient or the Laplacian. In some cases it was not possible to determine whether students had the concepts confused, or the names, or both:
\begin{quote}\textit{ `` $\boldsymbol{\nabla} \times\vec{A}$ is the vector product between the gradient and $\vec{A}$.'' }\end{quote}
Furthermore, about half of the students wrote down a formula from memory, some incorrectly; all of these are counted in the category `formula'. In the category `other' there were some correct statements that explained the link with Stokes' law, the divergence theorem or conservative fields. These students seem to have made connections with the integral form of Maxwell's equations.\par

On the whole, students seemed to feel these vector operators are a tool to evaluate something. A similar focus on evaluation was also seen in students' concept image of integration. In that particular case, students rather tried to evaluate an integral that was impossible to calculate than to describe it as an area under a curve or a sum of infinitesimal parts.~\cite{Doughty2014}\ This particular question does not elicit a structural understanding~\cite{Sfard1991,Tuminaro2004a}\ of the gradient, divergence and curl. The second part of the pretest is designed to investigate the students' ability to interpret graphical representations of vector fields and their skill at doing calculations in vector calculus. 

\subsubsection{Graphical interpretation of vector fields}
We gave our students a two-dimensional representation of four different vector fields and asked them to indicate where the divergence and curl are (non-)zero. The assignments can be found in Appendix \ref{app:pretest2}. The divergence is non-zero everywhere in Field~\ref{prefield1} and the curl is non-zero everywhere in Field~\ref{prefield2} and~\ref{prefield3}. The curl in Field~\ref{prefield4} clearly is zero everywhere, but determining the divergence is less straightforward. The field we sketched has $1/s$ dependence, so that the divergence is non-zero only at the center of the field. This could for example be the electric field of a charged wire pointing in the $z$-direction. However, if students saw an unspecified dependence on $s$ and stated that they could not decide whether the divergence was zero, we deemed their answer correct.\par

\begin{figure*}
\centering
{\includegraphics[angle=270, trim=5.5cm 2cm 5.5cm 2cm, clip=true, width=.8\linewidth]{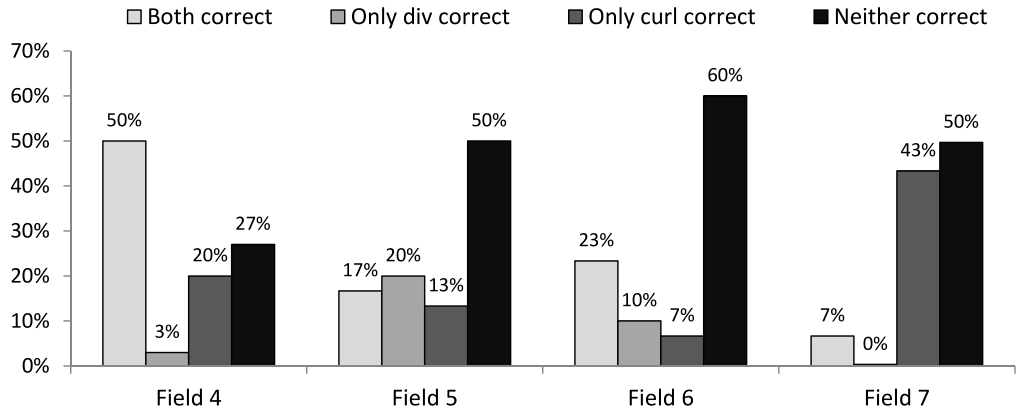}\caption{Results for the question about graphical representations of vector fields on the pretest ($N=30$).}\label{fig:pregraphical}}
\end{figure*}

First the students' answers were checked for correctness. Figure~\ref{fig:pregraphical}\ shows that our students have severe difficulties with these graphical representations. For Field~\ref{prefield1}, half of the students gave a correct answer, but for the other three vector fields less than one out of four figured out correctly where the divergence and curl are non-zero. In Field~\ref{prefield4}, just two students could determine that the divergence is non-zero only at the center of the field. About 30\% of the students made at least one statement that pointed toward the typical error~\cite{Pepper2012}\ of confusing the derivative of the field with its value (e.g. the derivative is zero when the field is zero). Our students were easily misdirected, and very inconsistent in their reasoning. Moreover, a significant number of students did not answer the question (about 20-30\% for the divergence and 30-40\% for the curl). It is likely that these students did not know how to solve these problems, since they did answer the other pretest questions.\par

Secondly, we looked at the strategy students used to obtain their answer. The prevalence of the strategies is shown in the second column of Table~\ref{tab:prestrategy}. We distinguish five categories:
\begin{itemize}
\item \emph{Concept based strategy}: This category includes explanations from students that show a good understanding of the underlying concepts. Their answers are based on drawings together with the definition and potentially some derived formulas that link the differential and integral form. Typically the change in flux per unit area is determined by drawing a small box around a point: if there is no net flux in the area bounded by the box, the divergence is zero as there is no source/sink in this area. To obtain the curl, a virtual paddle wheel is placed in the field. If it rotates, the curl is non-zero at that particular location.
\item	\emph{Formula based strategy}: The student mostly relies on ``the formula'' for divergence and curl, and uses the derivatives of $x$ and $y$ to get an answer. Some students even (try to) obtain an algebraic expression for the vector field and then apply the definition of the operator (typically in Cartesian coordinates).
\item	\emph{Description based strategy}: The students use a (correct/incorrect/incomplete) qualitative description of the divergence and curl to obtain an answer. Typically the student relies on the common English definition of the words `divergence' and `curl' and links this in naive way to the graphical representation of a vector field. This is illustrated by some examples for the third vector field:
\begin{itemize}
\item False descriptions:
\begin{quote}\textit{ `` $\boldsymbol{\nabla} \cdot \vec{A} \neq 0$ because the length of the arrows increases.'' }\end{quote}
\begin{quote}\textit{ `` $\boldsymbol{\nabla} \times\vec{A} = \vec{0}$ because the field is not rotating.'' }\end{quote}
\item More or less correct descriptions: 
\begin{quote}\textit{ `` $\boldsymbol{\nabla} \cdot \vec{A} = 0$ since nothing is added to the field anywhere.'' }\end{quote}
\begin{quote}\textit{ `` $\boldsymbol{\nabla} \times\vec{A} \neq \vec{0}$ because the field is rotating locally.'' }\end{quote}
\end{itemize}
It is of course possible that students who wrote down these descriptions have some conceptual insights as well, but their answers provided no evidence for this. 
\item	\emph{Unclear}: The reasoning is not explained or it is unclear.
\item	\emph{No answer}: The student did not answer the question.
\end{itemize}
Students generally used just one of these strategies to solve the question. However, some used a different approach to determine the divergence and the curl of a field.\par

\begin{table*}[htbp]
  \centering
  \caption{The prevalence and success rate of strategies students used to determine the divergence and curl of vector fields based on graphical representations used in the pretest $(N=30)$.}
	\begin{ruledtabular}
    \begin{tabular}{lccccc}
    \multicolumn{1}{c}{\multirow{2}[0]{*}{ {Divergence}}} & \multirow{2}[0]{*}{ {\# Students}} & \multicolumn{4}{c}{ {Success rate}} \\ 
		\cline{3-6}
    \multicolumn{1}{c}{} &       &  {Field~\ref{prefield1}} &  {Field~\ref{prefield2}} &  {Field~\ref{prefield3}} &  {Field~\ref{prefield4}} \\
    \hline
     {Concept based} & 3\% & 100\% & 100\% & 100\% & 100\% \\
     {Formula based} & 23\% & 71\% & 57\% & 57\% & 0\% \\
     {Description based} & 23\% & 71\% & 14\% & 14\% & 0\% \\
     {Unclear} & 30\% & 56\% & 44\% & 33\% & 11\% \\
     {No answer} & 20\% & -     & -     & -     & - \\
		\end{tabular}
		\end{ruledtabular}\\[0.3cm]
		\begin{ruledtabular}
		\begin{tabular}{lccccc}
    \multicolumn{1}{c}{\multirow{2}[0]{*}{ {Curl}}} & \multirow{2}[0]{*}{ {\# Students}} & \multicolumn{4}{c}{ {Success rate}} \\
		\cline{3-6}
    \multicolumn{1}{c}{} &       &  {Field~\ref{prefield1}} &  {Field~\ref{prefield2}} &  {Field~\ref{prefield3}} &  {Field~\ref{prefield4}} \\
    \hline
     {Concept based} & 3\% & 100\% & 100\% & 100\% & 100\% \\
     {Formula based} & 23\% & 100\% & 57\% & 71\% & 57\% \\
     {Description based} & 20\% & 100\% & 17\% & 33\% & 67\% \\
     {Unclear} & 23\% & 100\% & 29\% & 14\% & 71\% \\
     {No answer} & 30\% & -     & -     & -     & - \\
    \end{tabular}%
		\end{ruledtabular}
  \label{tab:prestrategy}%
\end{table*}%

In the last four columns of Table~\ref{tab:prestrategy}, the success rate for certain strategy is given. The student who used the concept based technique was very successful in determining the divergence and curl of the fields. The formula based technique is useful if the vector function can be found and calculations are carried out correctly, which may cause problems for complex cases (e.g. the fourth vector field). Students who used a description based strategy or give little or no explanation seem to have a low chance of being successful in determining the divergence and curl of a graphical representation of a vector field. There are some exceptions to these generalizations, like a student who determined the divergence and curl correctly for every field, but gave no explanation whatsoever. However, his answers were probably well considered, as he gave a fairly accurate description of the divergence and curl in the first part of the pretest:
\begin{quote}\textit{ ``The divergence of $\vec{A}$ is a scalar field that tells you how much is added to the vector field $\vec{A}$'' }\end{quote}
\begin{quote}\textit{ ``The curl of $\vec{A}$ is a vector field that tells you how strong and in which way the field $\vec{A}$ turns'' }\end{quote}
This leads us to believe that this particular student used his conceptual understanding of divergence and curl to tackle problems concerning the graphical representation of vector fields.\par

\subsubsection{Calculation of divergence and curl}
In the last set of questions on the pretest we asked our students to calculate the divergence and curl of three vector fields (see Appendix~\ref{app:pretest}). Two fields were given in Cartesian coordinates, the third in cylindrical coordinates. For each of the 6 calculations, we split answers into four categories: complete and correct calculations, calculations with minor mistakes or omissions (e.g. a forgotten minus sign or an expression is left unsimplified), calculations with major mistakes (e.g. an error in the use of the formula for div/curl, an error when taking the derivative or inappropriate use of unit vectors), or no answers. The results are presented in Table~\ref{tab:precalculate}.\par

\begin{table}[htbp]
  \centering
  \caption{Categorization of students' calculations of the divergence and curl of three vector fields in the pretest ($N=30$).}
	\begin{ruledtabular}
    \begin{tabular}{lcccccc}
          & \multicolumn{2}{c}{ {Exercise (a)}} & \multicolumn{2}{c}{ {Exercise (b)}} & \multicolumn{2}{c}{ {Exercise (c)}} \\
					\cline{2-3}\cline{4-5}\cline{6-7}
          &  {Div} &  {Curl} &  {Div} &  {Curl} &  {Div} &  {Curl} \\
		\hline
     {Correct} & 60\% & 53\% & 23\% & 43\% & 60\% & 57\% \\
     {Minor error} & 0\% & 20\% & 33\% & 13\% & 7\% & 3\% \\
		 {Major error} & 33\% & 16\% & 33\% & 3\% & 7\% & 7\% \\
		 {No answer} & 7\% & 10\% & 10\% & 40\% & 27\% & 33\% \\
    \end{tabular}%
		\end{ruledtabular}
  \label{tab:precalculate}%
\end{table}%

Exercise (a) required students to make a quite straightforward calculation. Nevertheless, only 60\% of the students were able to calculate the divergence correctly, and allowing for minor errors about three-quarters calculated the curl correctly. One student calculated the Laplacian instead of the curl. He did not give an answer for the other parts.\par

Exercise (b) was more difficult than the first one, since some challenging algebra is required to evaluate the expression in Cartesian coordinates. This explains the higher number of students who make minor errors. When allowing for minor errors, about 60\% of students gave correct answers. These are mostly students that could also correctly calculate the divergence and curl in the first exercise. Two students converted the equation to polar coordinates and then calculated the divergence and curl. One of them knew the formula for the divergence in 2 dimensions\footnote{This actually is the formula for the divergence in cylindrical coordinates in 3 dimensions, but with a $z$ component equal to zero.} (only the formulas for 3 dimensions were given), the other student made a mistake at this point. Both of them noticed that the curl is zero without doing any calculation at all. Two other students did not calculate the curl in this part, because they argued that a vector product is only defined in 3 dimensions. It may not have occurred to them that the vector field could be considered three-dimensional with zero $z$-component. While almost all students attempted to calculate the divergence, over one-third of the students did not attempt to calculate the curl. They may have been discouraged by difficulties they had when calculating the divergence.\par

The calculation required in Exercise (c) is as straightforward as that of Exercise (a), and a similar fraction of students calculated the divergence and curl correctly. Again, this is more or less the same group of students that could do the calculations in the first two exercises. However, many more students did not give an answer at all. It is unlikely that they did not know how to calculate the divergence and curl in cylindrical coordinates, since expressions were given to them.\par

In general, we can see that approximately 60\% of the students are able to calculate the divergence and the curl of given vector fields, independent of the level of difficulty and the coordinate system used, if we allow minor errors. The major errors can be classified in three subcategories: 10 times an error was made when taking the derivative, 11 students used the expressions incorrectly and in 10 cases unit vectors were used inappropriately (e.g. unit vectors were appended to terms in the divergence of a vector field). Of course a single student could make multiple errors during one calculation. Concerning the use of unit vectors, it was striking to see that students used them very inconsistently in the pretest.\par

When we compare the results of the calculations to the number of correct answers in the graphical representation question, the prior knowledge of these students clearly shows. They seem to have some difficulties calculating the divergence or curl, but struggle much more with exercises that ask for more insight. This confirms what we observed in the concept image of the students: most of our students lack a conceptual understanding of the divergence and curl, and focus on evaluation.

\subsection{Post-test}
After instruction up to Chapter 7 in Griffiths' textbook~\cite{Griffiths2012}, we gave the students a post-test questionnaire (see Appendix~\ref{app:posttest}). It comprises two questions concerning graphical interpretation of vector fields (one with and one without physics context), two questions where students have to calculate the divergence and curl after imposing a condition and two conceptual questions in which they had to use the differential form of Maxwell's equations to interpret a series of physical situations. Therefore the questions on the post-test correspond to the last three research questions in Section~\ref{sec:design}. We did not include a question that aims to examine the concept image of the divergence, curl and gradient because we wanted to exclude the possibility of a retest effect and to limit the workload for the students. Since the number of attendants in the non-mandatory lecture dropped over the semester, only 19 students filled in the post-test. All of these participants also took the pretest. Based on the pretest data the population of students that took the post-test is equivalent to the population of students that took the pretest.

\subsubsection{Graphical interpretation of (EM) vector fields}
\begin{figure*}
\centering
{\includegraphics[angle=270, trim=5.5cm 2cm 5.4cm 2cm, clip=true, width=.8\linewidth]{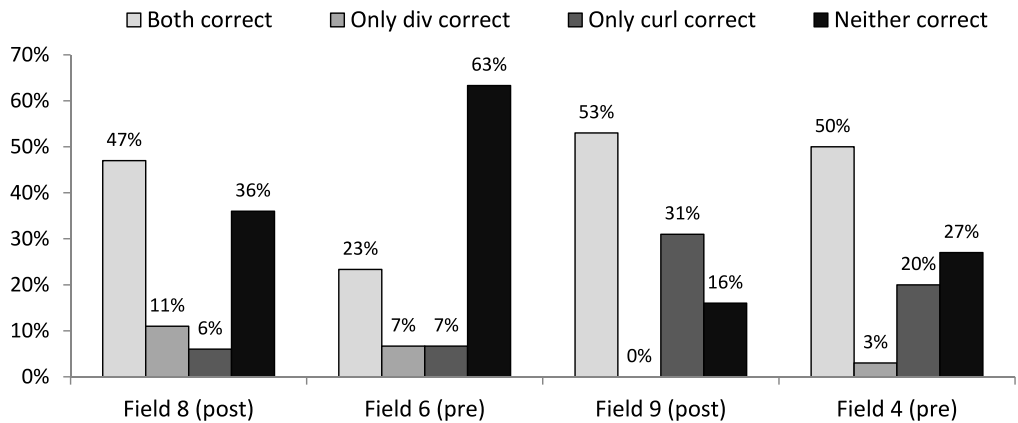}\caption{Results for the context-free graphical representation of vector fields on the post-test $(N=19)$ compared to the results on the pretest $(N=30)$. The first vector field on the post-test (Field~\ref{postfield1}) should be compared to Field~\ref{prefield3} on the pretest, and Field~\ref{postfield2} to Field~\ref{prefield1}.}\label{fig:postgraphical}}
\end{figure*}

\paragraph{Vector fields without physical context}~\\
This question is similar to the Question 1 of part 2 on the pretest. Field~\ref{postfield1} and~\ref{postfield2} on the post-test are analogous to Field~\ref{prefield3} and~\ref{prefield1} on the pretest respectively (however, the `view' is changed a bit). The results are summarized in Figure~\ref{fig:postgraphical}\ and are compared to the answers on the pretest.\par

It seems that students did better with the first vector field of the post-test, but made a more or less equal number of mistakes when interpreting the second vector field. This was analyzed more profoundly by looking at how many students' answers improved (incorrect at pretest; correct in post-test), disimproved (correct at pretest; incorrect at post-test) and stayed the same. For the first field, three students could correctly determine divergence and curl in the post-test, but not in the pretest. Not a single student made a `new' mistake. For the second field, four students improved their answers, but two went from answering correctly to answering incorrectly. This means that most students stick to their answers: there is a slight increase in correct answers, but still only about 50\% can determine the divergence and curl from a graphical representation of a simple vector field. Note that these percentages are similar to the results that Singh and Maries found when testing their graduate students (before instruction).\cite{Singh2013} \par

However, four students made incorrect statements like
\begin{quote}\textit{``The divergence is zero in the $x$ direction, but not in the $y$ direction.''}\end{quote}
This kind of reasoning was not seen in the graphical pretest questions, though it did emerge in the calculational pretest questions. We intend to explore this issue further in the future; it illustrates in any case that many students still struggle with divergence.\par

Looking at the strategies the students used to determine the divergence and curl (Table~\ref{tab:poststrategy}) we see an increase in concept based reasoning and formula based reasoning. The number of unclear answers decreased and every student at least tried to give an answer this time. This effect might be due to instruction: in lectures they were told that it is possible to use a paddle wheel to determine the curl, for example. However, students were not asked to use this idea in any exercises, which may explain the many errors they made. The students did a lot of calculations during the tutorial sessions and the fields are fairly straightforward, which may explain that formula based reasoning is more popular and effective in the post-test.\par

\begin{table}[htbp]
  \centering
  \caption{The prevalence and success rate of strategies students used to determine the divergence and curl of vector fields based on graphical representations used in the post-test $(N=19)$.}
    \begin{ruledtabular}
		\begin{tabular}{lccc}
    \multicolumn{1}{c}{\multirow{2}[0]{*}{ {Divergence}}} & \multirow{2}[0]{*}{ {\# Students}} & \multicolumn{2}{c}{ {Success rate}} \\
		\cline{3-4}
    \multicolumn{1}{c}{} &       &  {Field~\ref{postfield1}} &  {Field~\ref{postfield2}} \\
    \hline
     {Concept based} & 16\% & 67\% & 67\% \\
     {Formula based} & 37\% & 100\% & 100\% \\
     {Description based} & 26\% & 0\% & 20\% \\
     {Unclear} & 21\% & 0\% & 0\% \\
     {No answer} & 0\% & -     & - \\
		\end{tabular}
		\end{ruledtabular}\\[0.3cm]
		\begin{ruledtabular}
		\begin{tabular}{lccc}
    \multicolumn{1}{c}{\multirow{2}[0]{*}{ {Curl}}} & \multirow{2}[0]{*}{ {\# Students}} & \multicolumn{2}{c}{ {Success rate}} \\
		\cline{3-4}
    \multicolumn{1}{c}{} &       &  {Field~\ref{postfield1}} &  {Field~\ref{postfield2}} \\
    \hline
 {Concept based} & 21\% & 50\% & 50\% \\
     {Formula based} & 32\% & 100\% & 100\% \\
     {Description based} & 16\% & 0\% & 33\% \\
     {Unclear} & 32\% & 17\% & 17\% \\
     {No answer} & 0\% & - & - \\
		\end{tabular}%
		\end{ruledtabular}
  \label{tab:poststrategy}%
\end{table}%

\begin{figure}[h]
\centering
{\includegraphics[angle=270, trim=3.5cm 2.5cm 3cm 2.5cm, clip=true, width=\linewidth]{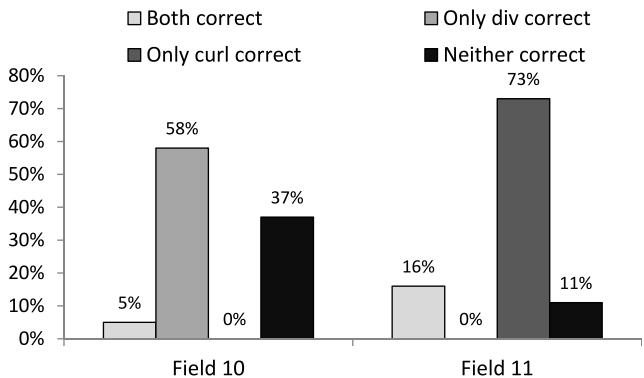}\caption{Percentage of correct answers for the graphical representation of electromagnetic fields in the post-test $(N=19)$.}\label{fig:graphicalem}}
\end{figure}

~\\ \paragraph{Electromagnetic fields}~\\
The post-test questions on graphical representations of electromagnetic fields ask similar questions in context. Furthermore, the fields are a bit more `difficult' in the sense that they have a cylindrical instead of a Cartesian symmetry. The number of correct answers is very small: only a few students could correctly determine both the divergence and the curl (Figure~\ref{fig:graphicalem}).\par

\begin{table}[htbp]
  \centering
  \caption{Strategies students used to interpret the divergence and curl of graphical representations of electromagnetic fields in the post-test $(N=19)$.}
    \begin{ruledtabular}
		\begin{tabular}{lcccc}
          \multicolumn{1}{c}{\multirow{2}[0]{*}{Prevalence}} & \multicolumn{2}{c}{ {Field~\ref{postfield3}}} & \multicolumn{2}{c}{ {Field~\ref{postfield4}}} \\
					\cline{2-3}\cline{4-5}
          &  {Div} &  {Curl} &  {Div} &  {Curl} \\
		\hline
     {Concept based} & 5\% & 16\% & 0\% & 0\% \\
     {Physics based} & 42\% & 11\% & 21\% & 26\% \\
     {Formula based} & 16\% & 11\% & 5\% & 0\% \\
     {Description based} & 16\% & 16\% & 11\% & 16\% \\
     {Unclear} & 26\%  & 42\% & 53\% & 58\%  \\
     {No answer} & 0\% & 5\% & 11\% & 5\% \\
    \end{tabular}
		\end{ruledtabular}\\[0.3cm]
		\begin{ruledtabular}
		\begin{tabular}{lcccc}
         \multicolumn{1}{c}{\multirow{2}[0]{*}{Success rate}} & \multicolumn{2}{c}{ {Field~\ref{postfield3}}} & \multicolumn{2}{c}{ {Field~\ref{postfield4}}} \\
					\cline{2-3}\cline{4-5}
          &  {Div} &  {Curl} &  {Div} &  {Curl} \\
		\hline
     {Concept based} & 100\% & 0\% & -     & - \\
     {Physics based} & 100\% & 50\% & 50\% & 100\% \\
     {Formula based} & 67\% & 0\% & 0\% & - \\
     {Description based} & 67\% & 0\% & 0\% & 100\% \\
     {Unclear} & 17\% & 0\% & 10\% & 91\%  \\
    \end{tabular}%
		\end{ruledtabular}
  \label{tab:strategyem}%
\end{table}%

To solve this question, students could use the same strategies as before, but could also use Maxwell's equations (physics based reasoning). Because some students used such an argument to confirm their answer based on another strategy, it is now possible that they are entered in multiple categories. The answers are summarized in Table~\ref{tab:strategyem}. The first part shows the percentage of students that used a certain approach. Some students used a generic approach together with a physics based strategy, so the total exceeds 100\%. The second part of the table shows the success rate when students use a certain strategy.\par

Students tend to change their strategy between questions (this is why both questions are treated separately) and often fail to correctly apply the strategies they use. They seem especially unsure about the use of Maxwell's equations. To illustrate this, we analyze the answer of a student for the magnetic field question (Field~\ref{postfield3}):
\begin{quote} Concerning the divergence, this student states correctly it is zero: \textit{``$\boldsymbol{\nabla} \cdot \vec{B}=0 \rightarrow$ there is no magnetic monopole''}, which is indeed always true. When looking at the curl, he writes \textit{``$\boldsymbol{\nabla} \times \vec{B}= \mu_0I$''}, but then crosses out the right hand side, and simply writes \textit{``non-zero''}. In the end, it is unclear what argument he used to obtain this incorrect result.\end{quote}
Many students did similar things: they tried to use Amp\`{e}re's law, but failed to apply it correctly. They did not appear to understand that the curl of a magnetic field is only non-zero where a current flows, and that it varies from point to point. Furthermore, they misinterpreted the use of a paddle wheel: they seemed to think it rotates everywhere, but it does so only in the center of the field (everywhere else it translates in a circle around the current carrying wire). Similar mistakes were made in the case of the electric field (Field~\ref{postfield4}): no more than three students understood that the divergence is non-zero only where charges are present. The most occurring mistake (37\%) was that students thought the divergence is nonzero everywhere because of the appearance of the field:
\begin{quote}\textit{``All arrows point towards a certain point, so the divergence is zero nowhere.''}\end{quote}
Formula based reasoning is less effective here, because the students struggle with the use of cylindrical coordinates or try to set up the equation of the field in Cartesian coordinates.\par

\subsubsection{Calculation of divergence and curl of EM fields}
These questions are intended to check if students can do calculations in an electromagnetism context, which takes the form of imposing a simple condition. To determine which field could be a magnetic field, students should check if $\boldsymbol{\nabla} \cdot \vec{B}=0$ applies; to check whether a field could be an electrostatic field, they need to verify that $\boldsymbol{\nabla} \times \vec{E}=\vec{0}$.\par
Recognizing and imposing this condition proved to be problematic for our students: 74\% (14) students were able to do this for the magnetic field and only 53\% (10) for the electric field. One student did not give answers to any of these questions. Others used qualitative reasoning that contained something about the radial or $z$-dependence:
\begin{quote}\textit{``A magnetic field spreads radially outward from its point of origin. The first one doesn't do that because it is in Cartesian coordinates.''}\end{quote}
A few students calculated the divergence of the potentially electrostatic fields, but then struggled to interpret the result.\par

Almost all students who obtained the correct condition calculated the divergence and curl correctly. Some even did not need to calculate a full expression, but could determine whether the divergence and curl were (non-)zero by sight. Despite the observation that some students still made errors, a slight progression could be noticed concerning the ability to perform calculations. This may be explained by the huge emphasis on calculations in Griffiths' textbook~\cite{Griffiths2012}\ and the exercise sessions.\par

\subsubsection{Conceptual understanding of Maxwell's equations}
The last set of post-test questions intends to investigate students' conceptual understanding of and insight into Maxwell's equations in differential form. To this end they had to determine whether the curl and divergence are zero or not in four electric and five magnetic fields. \footnote{After doing the analysis and having some fruitful discussions, we decided that a fifth electric field was described in an ambiguous way, and the physics was more involved than intended. Therefore we left this situation out of the analysis.} When Maxwell's equations in differential form are applied correctly in every situation, one obtains the answers: $\boldsymbol{\nabla} \cdot \vec{E}=0$ always except for the first situation, $\boldsymbol{\nabla} \times\vec{E}=\vec{0}$ in situation a, c, and d, $\boldsymbol{\nabla} \cdot \vec{B}=0$ always and $\boldsymbol{\nabla} \times\vec{B}=\vec{0}$ in the last two situations. We did not give the students a list of Maxwell's equations to avoid pointing students in a particular direction. In our opinion students who understand these laws will be able to reproduce at least the causality between the fields and sources. However, as shown in Table~\ref{tab:conceptem}, students encountered tremendous difficulties in answering this question.\par

\begin{table}[htbp]
  \centering
  \caption{The percentage of students who could correctly determine whether the divergence and curl of the electric (magnetic) field described in a situation is zero or not $(N=19)$.}
    \begin{ruledtabular}
    \begin{tabular}{lcc}
    \multicolumn{1}{c}{ {Electric field}} &  {$\boldsymbol{\nabla} \cdot \vec{E}$} &  {$\boldsymbol{\nabla} \times\vec{E}$}\\
		\hline
     {Situation a} & 47\% & 95\%  \\
     {Situation b} & 16\% & 53\%  \\
     {Situation c} & 26\%  & 74\%  \\
     {Situation d} & 95\%  & 95\%  \\
		\end{tabular}
		\end{ruledtabular}\\[0.3cm]
		\begin{ruledtabular}
		\begin{tabular}{lcc}
    \multicolumn{1}{c}{ {Magnetic field}} &  {$\boldsymbol{\nabla} \cdot \vec{B}$} &  {$\boldsymbol{\nabla} \times\vec{B}$} \\
		\hline
     {Situation a} & 63\%  & 79\%  \\
     {Situation b} & 74\%  & 79\%  \\
     {Situation c} & 74\%  & 42\%  \\
     {Situation d} & 84\%  & 74\%  \\
     {Situation e} & 84\%  & 68\%  \\
    \end{tabular}%
		\end{ruledtabular}
  \label{tab:conceptem}%
\end{table}%

Only one student did not make a single mistake, while all other 18 students made at least three errors. We could not find a correlation between the errors, but some patterns did emerge:
\begin{itemize}
\item Gauss' law, which states that the divergence of the electric field is non-zero only where charge is present, or more conceptually still, that the source (sink) of an electric field is a positive (negative) electric charge, elicited most errors. Only one student could correctly determine where the divergence is non-zero for all five situations. This confirms some of the findings from the University of Colorado: students have difficulties applying the divergence in an electromagnetic context.~\cite{Pepper2012}
\item At least one mistake was made in the application of Faraday's law by 53\% (10) students. Nine students did not appear to know that the curl of an electric field is non-zero when there is a changing magnetic field, as stated explicitly in situation b.
\item Despite the elaborate discussion on the non-existence of magnetic monopoles during instruction, 53\% (10) students did not check every box under $\boldsymbol{\nabla} \cdot \vec{B}$. This is alarming, since it is a truly elementary law that is easy to apply.
\item Only 21\% (4) students were able to correctly evaluate Amp\`{e}re-Maxwell's law in every situation. Strikingly, fewer than half of the students could interpret situation c, which is the classic textbook example to show Maxwell's correction to Amp\`{e}re's law. It was both discussed during the lectures and in Griffiths' textbook.~\cite{Griffiths2012}
\end{itemize}
These results show that our students did not profoundly understand Maxwell's equations in differential form.\par

\section{Conclusions and implications for teaching}\label{sec:conclusions}
We have investigated students' understanding of divergence and curl in mathematical and physical contexts. Concerning their initial concept image, we found that they focused on evaluation, and appeared to pay little attention to the conceptual meaning of the vector operators. Furthermore, their conceptual descriptions often were incomplete and contained incorrect information. Some students were confused about the vector or scalar character of the operators, and used incorrect terminology.\par
Interpreting graphical representations of vector fields is a difficult exercise for students. Even after instruction only half of the students were able to determine where the divergence and curl of a simple vector field are (non)-zero. When more complex and realistic electromagnetic fields had to be considered, only a few students succeeded in solving the question correctly. Moreover, many students used various strategies inconsistently. This suggests they lack a structural understanding of the mathematical concepts, and on top of this they are unable to use their acquired skills in a physical context.\par
Since we (and many others~\cite{Gire2012,Singh2013,Pepper2012}) believe that these graphical representations are helpful when trying to conceptualize the abstract mathematical structures in vector calculus, we think it would be advisable to put more effort into this kind of exercises in both physics and mathematics instruction. In our opinion it would help students to understand the physical meaning of Maxwell's equations, which has applications beyond E\&M -- e.g. in subsequent problems concerning electromagnetic radiation, gauge theory and the introduction to special relativity.\par
In Griffiths' textbook~\cite{Griffiths2012}\ a lot of exercises focus on complex calculations. We found that students are reasonably comfortable with the required algebra, but have problems when they need to interpret the context of a calculation. One out of four students was unable to come up with the condition a vector field should satisfy in order to be a realistic magnetic field, and only half of them knew this condition for an electrostatic field. Since understanding and explaining electrodynamic phenomena is one of the main objectives of this course, we suggest more attention should be paid to the interpretation of the equations and setting up the problem at the expense of doing algebraic manipulations.\par
When we investigated the competences students show when they encounter situations that can be solved using Maxwell's equations in differential form, we observed that students had tremendous difficulties with the application of all four laws. This calls for an instruction that puts more effort in linking mathematics and physics and uses a more qualitative approach. Some great ideas can be found in the article by Huang et al \cite{Huang2013}, although we think even more graphical and conceptual examples are needed in order to fully show students the power and usefulness of Maxwell's equations in differential form.\par
In future work we are planning to conduct student interviews to gain more insight in the thinking process of students when they solve problems linked to the differential form of Maxwell's equations. This will help us understand how graphical representation and a better structural understanding of the mathematical concepts can help students to apply their skills in a physical context. At a later stage the results of these interviews will be used to create new or improved questions on the pre- and post-test, and to iteratively design a tutorial that aims to help students understanding Maxwell's equations in differential form.

\begin{acknowledgments}
We gratefully acknowledge fruitful discussions with Charles Baily and thank him for his valuable comments on the paper. We greatly appreciate the cooperation with Wojciech De Roeck and the students who made this study possible.

\end{acknowledgments}

\appendix

\section{Pretest questions} \label{app:pretest}
The first part was given to the students without any expressions for div, grad and curl. After they finished and turned in Part 1, they were given Part 2 which contained the expressions given in Appendix~\ref{app:formulas}. Some space was left blank for the students to answer after each question. Figures are displayed smaller than in the actual test.
\subsection*{Part 1} \label{app:pretest1}
\begin{enumerate}[leftmargin=*]
\item Interpret (i.e. write down everything you think of when you see) the following operations.
	\begin{enumerate}
	\item $\displaystyle  \boldsymbol{\nabla} A $
	\item $\displaystyle  \boldsymbol{\nabla} \cdot \vec{A} $
	\item $\displaystyle  \boldsymbol{\nabla} \times\vec{A} $
	\end{enumerate}
\end{enumerate}

\subsection*{Part 2} \label{app:pretest2}

\begin{enumerate}[leftmargin=*]
\item Indicate where the divergence and/or curl is (non)zero for the next vector fields in the $x,y$-plane. The $z$-component is zero everywhere. Explain and show your work.
\begin{figure}[H]
\centering
\begin{minipage}{.5\linewidth}
  \centering
	{\includegraphics[trim=1.5cm 0.5cm 1.5cm 0.5cm, clip=true, width=\linewidth]{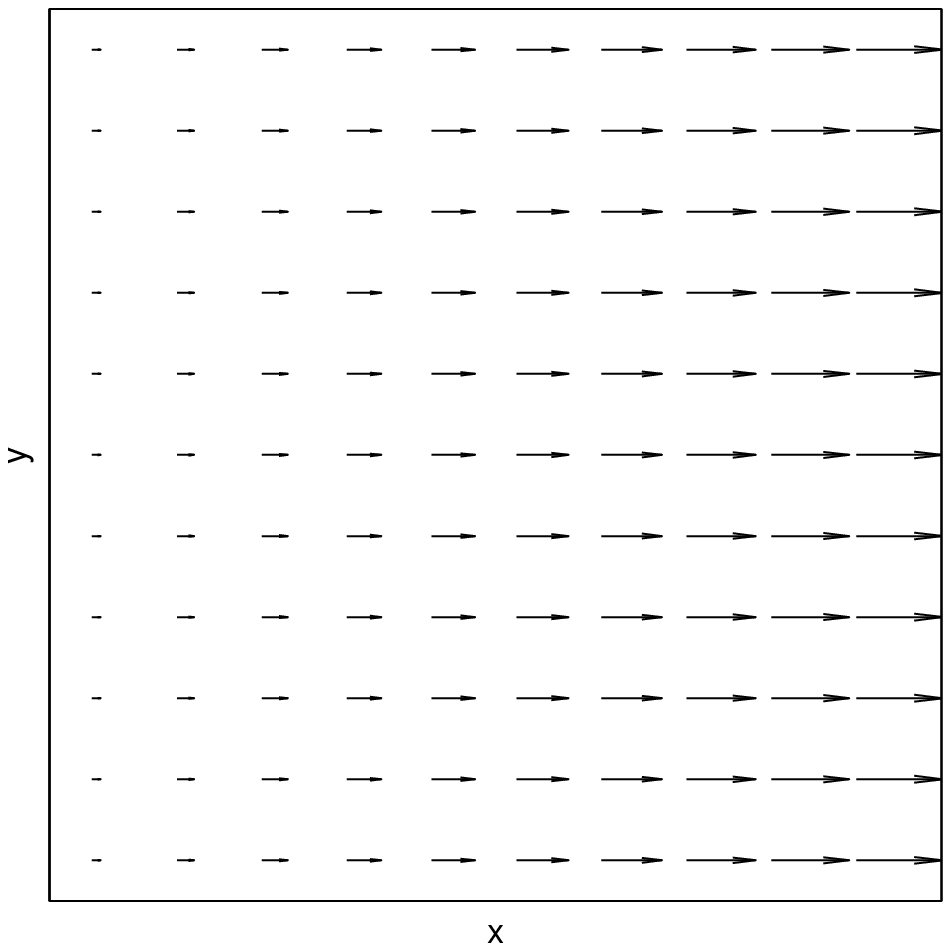}\caption{Pretest field 1}\label{prefield1}}
\end{minipage}%
\begin{minipage}{.5\linewidth}
  \centering
	{\includegraphics[trim=1.5cm 0.5cm 1.5cm 0.5cm, clip=true, width=\linewidth]{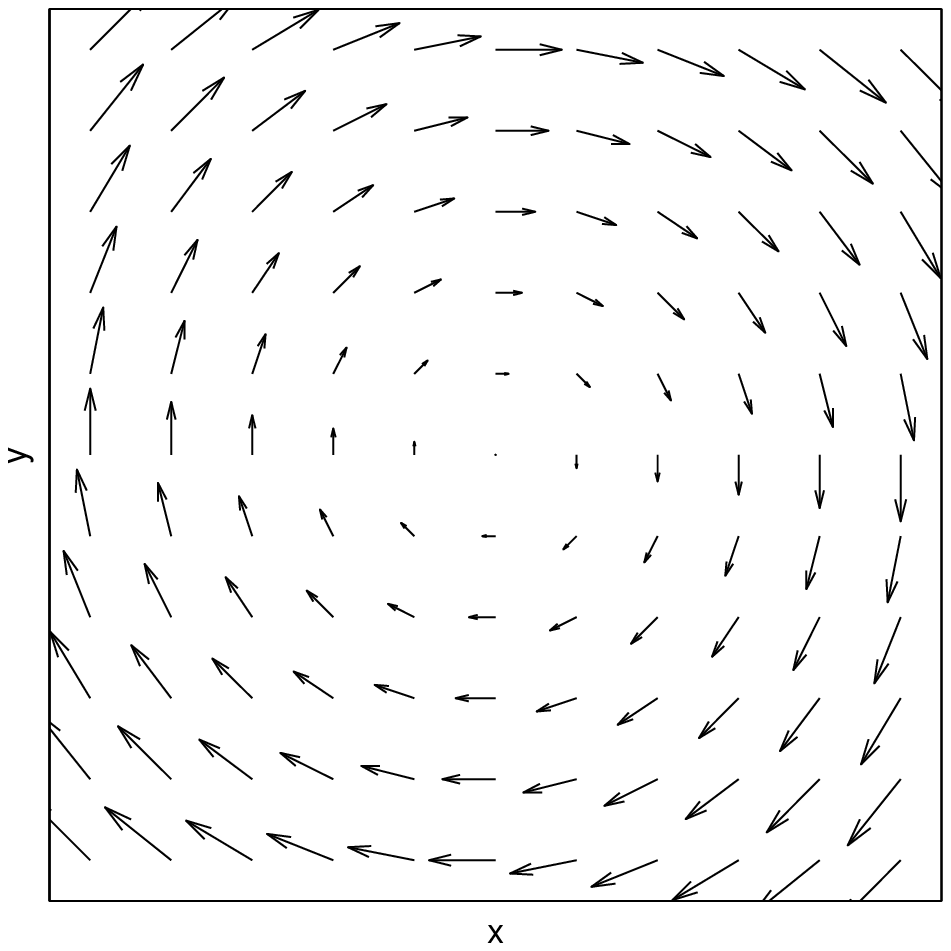}\caption{Pretest field 2}\label{prefield2}}
\end{minipage}
\end{figure}
\begin{figure}[H]
\centering
\begin{minipage}{.5\linewidth}
  \centering
	{\includegraphics[trim=1.5cm 0.5cm 1.5cm 0.5cm, clip=true, width=\linewidth]{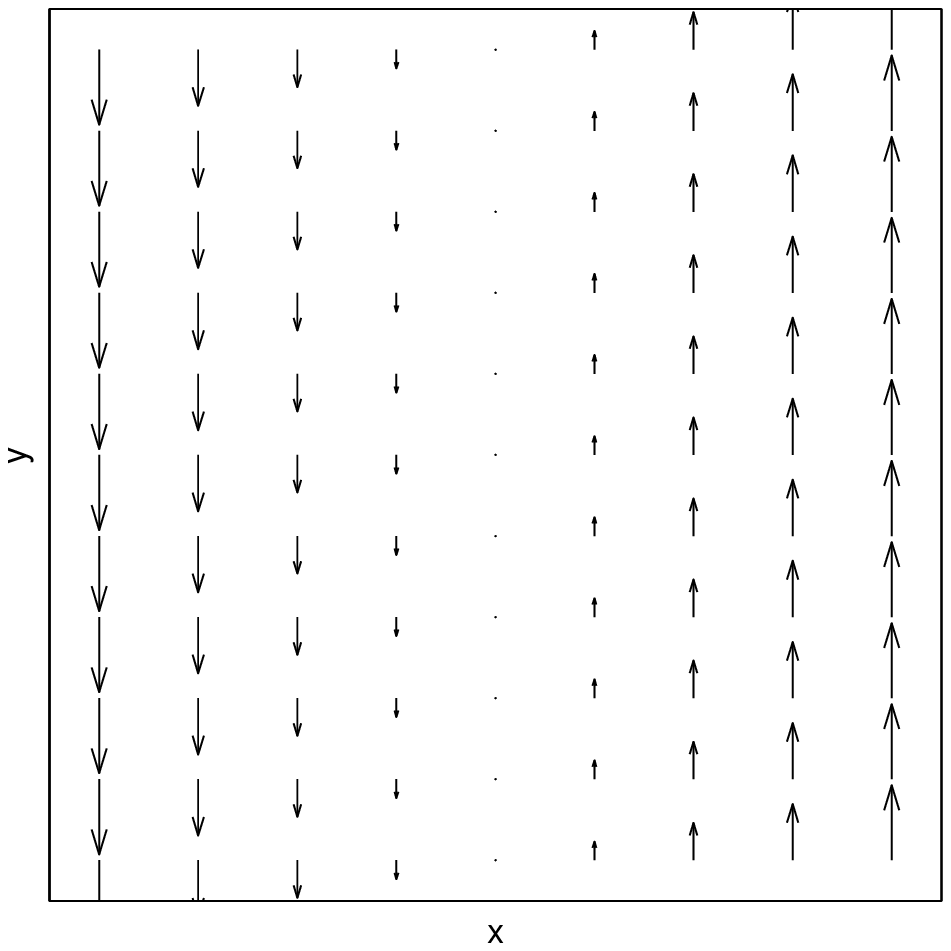}\caption{Pretest field 3}\label{prefield3}}
\end{minipage}%
\begin{minipage}{.5\linewidth}
  \centering
	{\includegraphics[trim=1.5cm 0.5cm 1.5cm 0.5cm, clip=true, width=\linewidth]{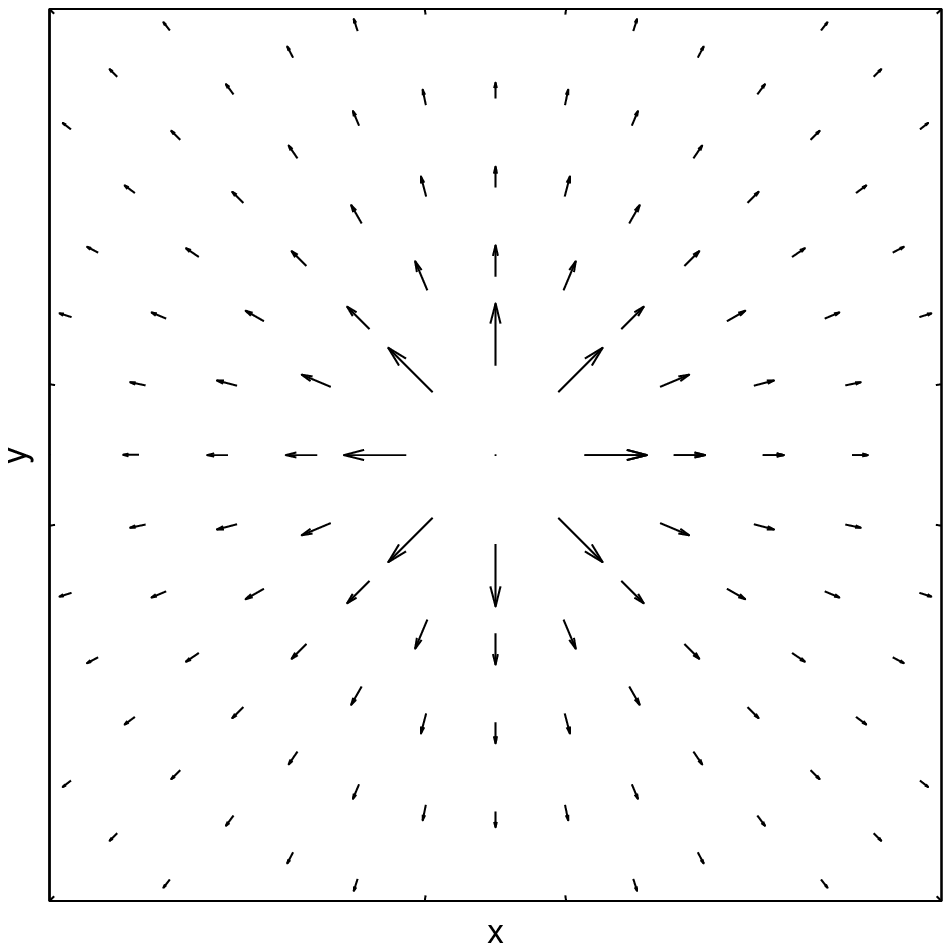}\caption{Pretest field 4}\label{prefield4}}
\end{minipage}
\end{figure}

\item Calculate the divergence and curl of the following vector fields.
	\begin{enumerate}
	\item $\displaystyle \vec{v}_a=  x^2 \hat{e}_x +  x \hat{e}_y - 2xz\hat{e}_z$ 
	\item $\displaystyle \vec{v}_b=  \frac{x\hat{e}_x +  y\hat{e}_y}{(x^2+y^2)^{3/2}} $ 
	\item $\displaystyle \vec{v}_c=  (r/2,r\theta,-z)$ (Hint: cylindrical coordinates) 
	\end{enumerate}
\end{enumerate}

\section{Post-test questions} \label{app:posttest}
The expressions of Appendix~\ref{app:formulas}\ were appended to these questions. Some space was left blank for the students to answer after each question. Figures are displayed smaller than in the actual test.
\begin{enumerate}[leftmargin=*]
\item Indicate where the divergence and/or curl is (non-)zero for the following vector fields in the $(x,y)$ plane. The $z$-component is zero everywhere. Explain and show your work.
\begin{figure}[H]
\centering
\begin{minipage}{.5\linewidth}
  \centering
	{\includegraphics[trim=1.5cm 0.5cm 1.5cm 0.5cm, clip=true, width=\linewidth]{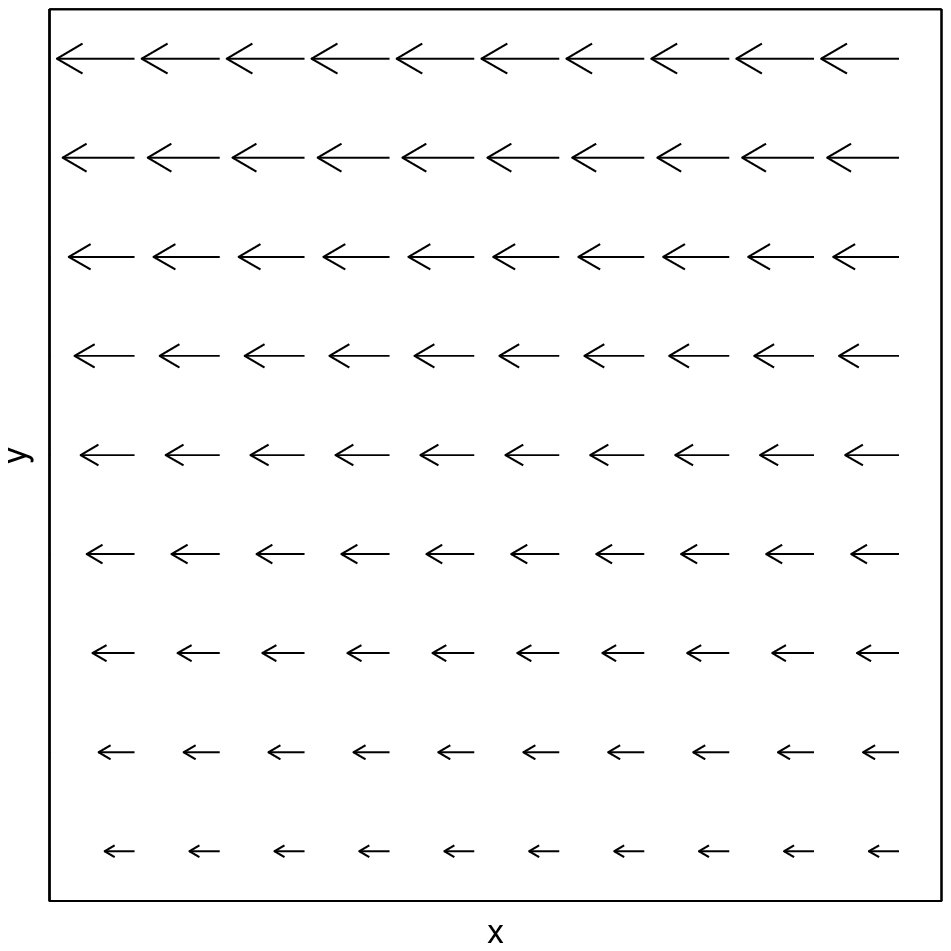}\caption{Post-test field 1}\label{postfield1}}
\end{minipage}%
\begin{minipage}{.5\linewidth}
  \centering
	{\includegraphics[trim=1.5cm 0.5cm 1.5cm 0.5cm, clip=true, width=\linewidth]{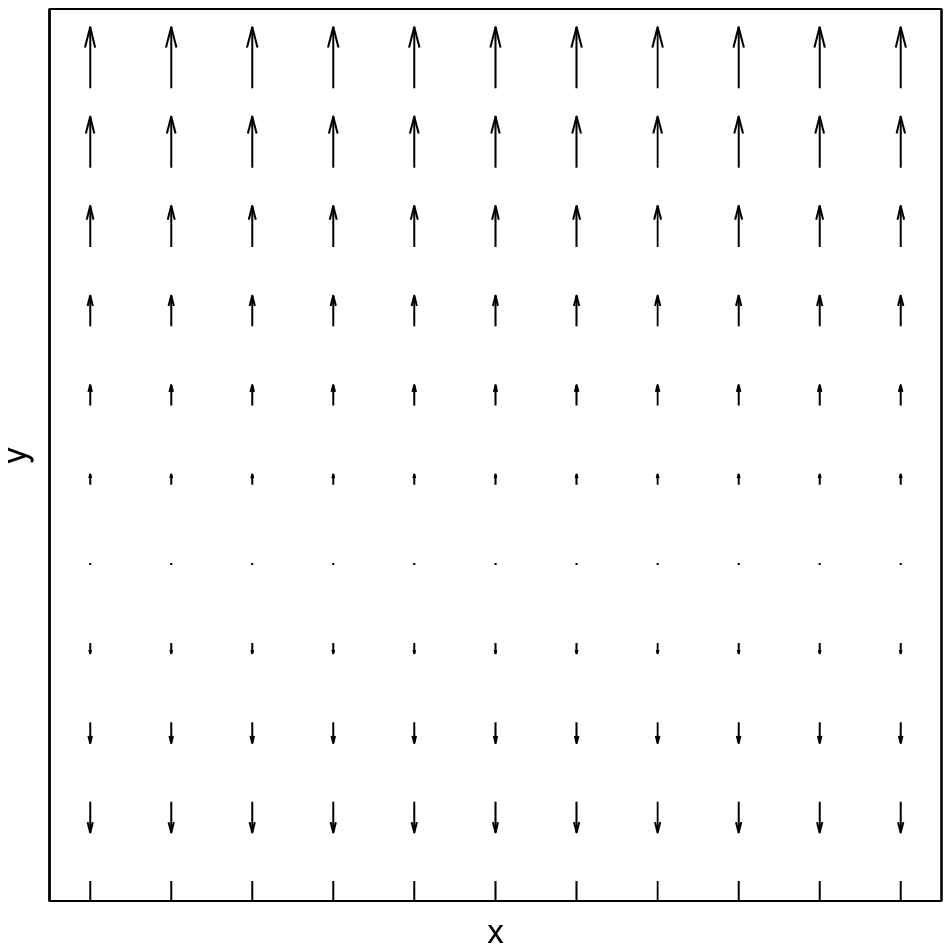}\caption{Post-test field 2}\label{postfield2}}
\end{minipage}
\end{figure}

\item For the following physical situations, explain where the divergence and/or curl of the field are (non-)zero. The $z$-component of the fields is zero everywhere. Show your work.
\begin{enumerate}
\item The magnetic field of an infinite current carrying wire along the $z$-axis.
\begin{figure}[H]
\centering
{\includegraphics[trim=1.5cm 0.5cm 1.5cm 0.5cm, clip=true, width=0.5\linewidth]{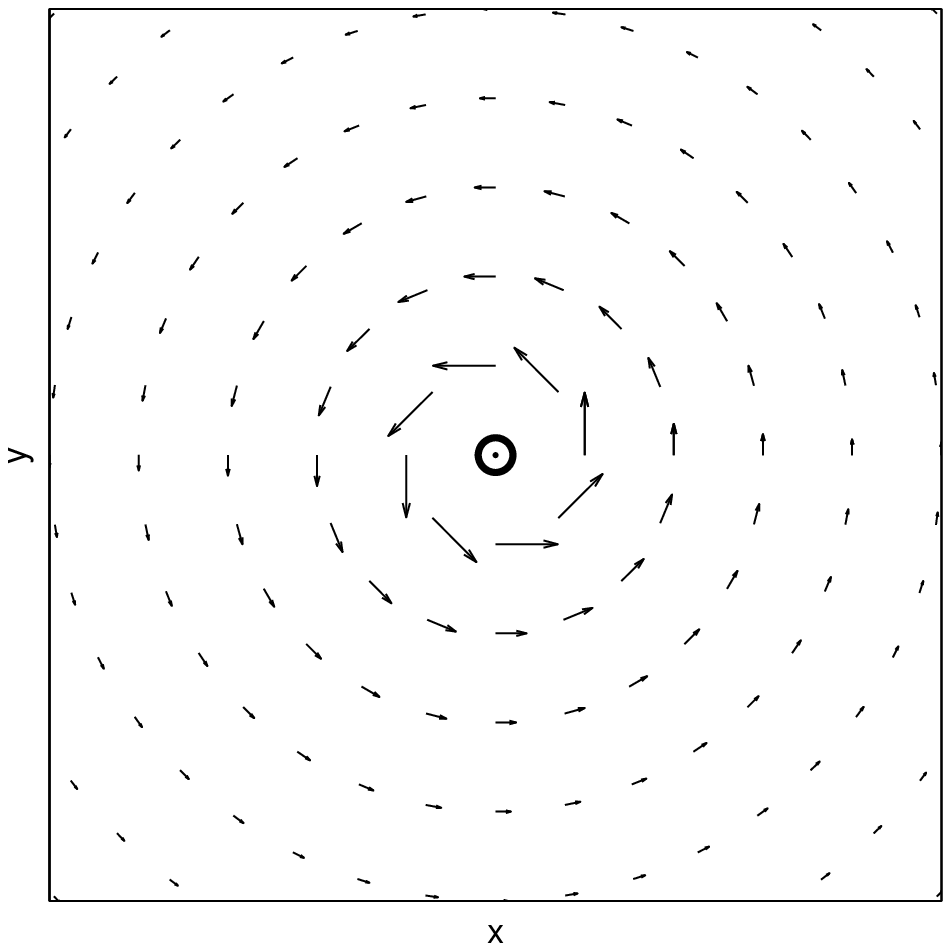}\caption{Post-test field 3}\label{postfield3}}
\end{figure}
\item The electric field of a charged infinitely long cylinder with radius $R$. In the figure, the cross-section in the $x,y$ plane is given.
\begin{figure}[H]
\centering
{\includegraphics[trim=1.5cm 0.5cm 1.5cm 0.5cm, clip=true, width=0.5\linewidth]{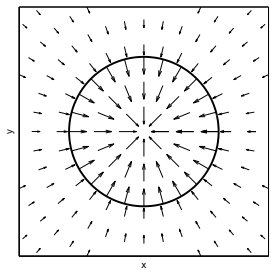}\caption{Post-test field 4}\label{postfield4}}
\end{figure}
\end{enumerate}

\item Which of these equations could represent a realistic magnetic field ($B_0$ is a constant with the appropriate units)? Explain. 
	\begin{enumerate}
	\item $\displaystyle \vec{B}_a = B_0[ 4xy   	    \hat{x} -y^2 	\hat{y} + (x-2yz) \hat{z}] $
	\item $\displaystyle \vec{B}_b = B_0[ \hat{r} + 4r^2 \hat{\theta} - 2\sin(\theta)\hat{\phi}] $
	\end{enumerate}
	
\item Which of these equations could represent a realistic static electric field ($E_0$ is a constant with the appropriate units)? Explain.
		\begin{enumerate}
		\item $\displaystyle \vec{E}_a= E_0 [ (z-x) \hat{x} +(z+x)   \hat{y} +x       \hat{z} ]$ 
		\item $\displaystyle \vec{E}_b= E_0 [ s(2+\sin^2{\phi}) \hat{s}+  s\sin{\phi}\cos{\phi}	\hat{\phi}+ 3z \hat{z}]$
		\end{enumerate}

\newcommand{\spatie}{\hspace{0.3cm}} 
\newcommand{\checkbox}{{\large\raisebox{-0.6ex}{$\Box$}\hspace{.6ex}}} 

\item Check the box(es) if $\boldsymbol{\nabla} \cdot \vec{E}$ and/or $\boldsymbol{\nabla} \times \vec{E}$ are equal to zero. 
\begin{table}[H]
  \centering
	\begin{enumerate}[leftmargin=*,itemsep=5pt,topsep=0pt,partopsep=0ex,parsep=0ex]
    \begin{tabular}{m{0.5\linewidth} m{0.2\linewidth} m{0.2\linewidth}}
							&				\multicolumn{1}{c}{\spatie $\boldsymbol{\nabla}\cdot \vec{E}=0$ \spatie}	&		\multicolumn{1}{c}{\spatie $\boldsymbol{\nabla}\times \vec{E} = \vec{0}$ \spatie}		\\
		\item The electric field at a distance $r<R$ from the center of a uniformly charged sphere with radius $R$.  &   \multicolumn{1}{c}{\spatie \checkbox \spatie}    & \multicolumn{1}{c}{\spatie \checkbox \spatie} \\
    \item The electric field generated by a changing magnetic field. &   \multicolumn{1}{c}{\spatie \checkbox \spatie}    & \multicolumn{1}{c}{\spatie \checkbox \spatie} \\
    \item The electric field at a distance $r$ from a pure electric dipole. &   \multicolumn{1}{c}{\spatie \checkbox \spatie}    & \multicolumn{1}{c}{\spatie \checkbox \spatie} \\
		\item The electric field inside a charged conductor.&   \multicolumn{1}{c}{\spatie \checkbox \spatie}    & \multicolumn{1}{c}{\spatie \checkbox \spatie} \\
    \end{tabular}
		\end{enumerate}
\end{table}

\item Check the box(es) if $\boldsymbol{\nabla}\cdot \vec{B}$ and/or $\boldsymbol{\nabla}\times \vec{B}$ are equal to zero.
\begin{table}[H]
  \centering
	\begin{enumerate}[leftmargin=*,itemsep=5pt,topsep=0pt,partopsep=0ex,parsep=0ex]
    \begin{tabular}{m{0.5\linewidth} m{0.2\linewidth} m{0.2\linewidth}}
							&				\multicolumn{1}{c}{\spatie $\boldsymbol{\nabla}\cdot \vec{B}=0$ \spatie}	&		\multicolumn{1}{c}{\spatie $\boldsymbol{\nabla}\times \vec{B} = \vec{0}$ \spatie}		\\
		\item The magnetic field generated by a changing electric field.   &   \multicolumn{1}{c}{\spatie \checkbox \spatie}    & \multicolumn{1}{c}{\spatie \checkbox \spatie} \\
      \item The magnetic field at a distance $r<R$ from the axis of a cylindrical conductor with radius $R$ carrying a steady current. &   \multicolumn{1}{c}{\spatie \checkbox \spatie}    & \multicolumn{1}{c}{\spatie \checkbox \spatie} \\
     \item The magnetic field between the plates of a charging capacitor. &   \multicolumn{1}{c}{\spatie \checkbox \spatie}    & \multicolumn{1}{c}{\spatie \checkbox \spatie} \\
		\item The magnetic field inside a solenoid with a steady current passing through it. &   \multicolumn{1}{c}{\spatie \checkbox \spatie}    & \multicolumn{1}{c}{\spatie \checkbox \spatie} \\
		\item The magnetic field at a distance $r$ of a large conducting plate carrying a steady surface current density $\vec{K}$.&   \multicolumn{1}{c}{\spatie \checkbox \spatie}    & \multicolumn{1}{c}{\spatie \checkbox \spatie} \\
    \end{tabular}
		\end{enumerate}
\end{table}

\end{enumerate}

\section{Formulas} \label{app:formulas}
This section contains the formulas that were handed to the students with the pretest (part 2) and post-test. In the pretest, these equations were given using the notation the students learned in their calculus courses. In the post-test we used Griffiths' notation.\cite{Griffiths2012} Only the latter are presented here.

\begin{itemize}
	\item $\displaystyle \boldsymbol{\nabla} \cdot \vec{v} = \frac{\partial v_x}{\partial x} + \frac{\partial v_y}{\partial y} + \frac{\partial v_z}{\partial z}$
	\item $\displaystyle \boldsymbol{\nabla} \times \vec{v} =  \left( \frac{\partial v_z}{\partial y} - \frac{\partial v_y}{\partial z}\right) \hat{x} + \left( \frac{\partial v_x}{\partial z} - \frac{\partial v_z}{\partial x}\right) \hat{y} + \left( \frac{\partial v_y}{\partial x} - \frac{\partial v_x}{\partial y}\right) \hat{z}$
	\end{itemize} \vfill
\begin{itemize}
	\item $\displaystyle \boldsymbol{\nabla} \cdot \vec{v} = \frac{1}{r^2} \frac{\partial}{\partial r} (r^2 v_r) + \frac{1}{r \sin \theta} \frac{\partial}{\partial \theta} (\sin \theta  v_\theta) + \frac{1}{r \sin \theta} \frac{\partial v_\phi}{\partial \phi}$
	\item $  \displaystyle \boldsymbol{\nabla} \times \vec{v} =   \frac{1}{r \sin \theta} \left[ \frac{\partial}{\partial \theta} (\sin \theta v_\phi ) - \frac{\partial v_\theta}{\partial \phi} \right] \hat{r}  + {\frac{1}{r} \left[ \frac{1}{\sin \theta} \frac{\partial v_r}{\partial \phi} - \frac{\partial}{\partial r} (r v_\phi) \right] \hat{\theta} + \frac{1}{r} \left[ \frac{\partial}{\partial r} (r v_\theta) - \frac{\partial v_r}{\partial \theta} \right] \hat{\phi} }$ 
	\end{itemize} \vfill
\begin{itemize}
	\item $\displaystyle \boldsymbol{\nabla} \cdot \vec{v} = \frac{1}{s} \frac{\partial}{\partial s} (s v_s) + \frac{1}{s} \frac{\partial v_\phi}{\partial \phi} + \frac{\partial v_z}{\partial z}$
	\item $\displaystyle \boldsymbol{\nabla} \times \vec{v} = \left[ \frac{1}{s} \frac{\partial v_z}{\partial \phi} - \frac{\partial v_\phi}{\partial z} \right] \hat{s} + \left[ \frac{\partial v_s}{\partial z} - \frac{\partial v_z}{\partial s} \right] \hat{\phi} + \frac{1}{s} \left[ \frac{\partial}{\partial s} (s v_\phi) - \frac{\partial v_s}{\partial \phi} \right] \hat{z}$
\end{itemize}
\vfill

\input{Bibliography.bbl}

\end{document}

%% file: Bibliography.bbl
%